\documentstyle[12pt]{article}
\newcommand{\be}{\begin{eqnarray}}
\newcommand{\ee}{\end{eqnarray}}
\font\tenrm=cmr10
\font\tensl=cmsl10

\begin{document}
\baselineskip=16pt
\begin{center}
\tenrm
{\Large\bf Deep Inelastic
Scattering of Polarized Electrons}
\\
{\Large\bf off Polarized $^3$He:
Nuclear Effects and}
\\
{\Large\bf the Neutron Spin Structure Functions
\footnote[1]{
Invited talk at VI Workshop on
``Perspectives in Nuclear Physics at Intermediate Energies", ICTP, Trieste,
May 3-7, 1993 (World Scientific, Singapore). }}
\\
[0.8cm]
C. CIOFI degli ATTI\\
{\tensl Dipartimento di Fisica, Universit\`a di
Perugia,\\ and \\ INFN, Sezione di Perugia,
Via A. Pascoli, I-06100 Perugia, Italy}\\[0.5cm]
E. PACE\\
{\tensl Dipartimento di Fisica, Universit\`a di Roma ``Tor Vergata",\\
and \\INFN,
Sezione Tor Vergata, Via E. Carnevale, I-00173 Roma, Italy}
\\[0.5cm]
 G. SALM\`E
\\
{\tensl INFN, Sezione Sanit\`a, Viale Regina Elena 299, I-00161Roma, Italy}
\\[0.5 cm]
S. SCOPETTA
\\
{\tensl Dipartimento di Fisica, Universit\`a di
Perugia,\\ and \\ INFN, Sezione di Perugia,
Via A. Pascoli, I-06100 Perugia, Italy}
\\[.8cm]
\end{center}

\begin{abstract}
{\tenrm
Nuclear effects
in Deep Inelastic Scattering
of polarized electrons off polarized $^3$He
are analyzed in terms of a spin dependent spectral function
taking into account
$S'$ and $D$ waves in $^3$He, as well as Fermi motion and binding
effects. A simple and reliable equation
relating the neutron
and $^3$He spin structure functions
is proposed.
}
\end{abstract}

The spin structure functions (SSF) of the nucleon $g_1^N$ and
$g_2^N$ provide information on the spin distribution among the
nucleon partons and can allow
important tests of various models of hadron's structure \cite{gen}.
The proton SSF $g_1^p$ has been measured in
\cite{slac1}; the first data on the neutron SSF were recently obtained
by the E142 and SMC collaborations \cite{smc,slac2}; future
experiments \cite{herm} will
improve the knowledge
on $g_1^p$ and $g_1^n$ and will provide first measurements
of
$g_2^p$ and $g_2^n$.
The neutron SSF are obtained from the
spin
asymmetry
measured in Deep Inelastic Scattering (DIS) of longitudinally
polarized electrons off polarized nuclear targets, viz. $\vec {^2{\rm H}}$
and $\vec {^3{\rm He}}$. As is well known, the use
of $\vec{^3{\rm He}}$ targets, which will be considered in this talk,
is motivated by the observation that,
in the simplest picture of $^3$He
(all nucleons in $S$ wave),
protons have
opposite spins, so
that their
contribution to the asymmetry largely cancels out. However, such a cancellation
does not occur
if other components
of the three body wave function are considered;
moreover, the fact that electrons scatter
off nucleons having a certain momentum and energy distribution
may, in principle, limit the possibility to obtain information
on nucleon SSF from scattering
experiments on nuclear targets.
In Ref. \cite{pg}, the question has been quantitatively discussed as to whether
and to what extent the extraction of $g_1^n$
from the asymmetry of the process
$ {^3\vec{\rm He}}({\vec e}, e')$X could be
hindered by nuclear effects arising
from small wave function components of $^3$He, as well as from
Fermi motion and binding
effects on DIS.
In this talk, the approach and main conclusions of Ref. \cite{pg} will be
discussed, and
the first
theoretical predictions on the SSF
$g_2^3$ and $g_2^N$ will be presented. The basic nuclear ingredient
used in our calculations is
the spin dependent spectral
function of $^3$He \cite{cps1}, which allows one to take into account
at the same time Fermi motion and binding corrections
(unlike Ref. \cite{wol} where the $\vec {^3{\rm He}}$
asymmetry has been calculated taking into
account $S'$ and $D$ waves but considering only Fermi motion and omitting
$Q^2$ dependent terms);
moreover,
our work is based on a
recent, improved theoretical description of inclusive scattering
of polarized electrons by polarized nuclei
\cite{sa}.

For in\-clu\-si\-ve scattering
of longitudinally polarized electrons off a
polarized $J={1\over2}$ target with atomic
weight $A$, the longitudinal asymmetry
reads as follows:
\be
A_{||}={\sigma_{\uparrow \uparrow} - \sigma_{\uparrow \downarrow}
\over \sigma_{\uparrow \uparrow} + \sigma_{\uparrow \downarrow}}=
2x [1+R(x,Q^2)]{g_1^A(x,Q^2)-{Q^2 \over \nu (\epsilon_1 + \epsilon_2
\cos \theta)}
g_2^A(x,Q^2)\over F_2^A(x,Q^2)} \equiv A_{\vec A} \label{as}
\ee
where $\sigma_{\uparrow \uparrow (\uparrow \downarrow)}$ is the
differential cross section corresponding to the target spin parallel
(antiparallel) to the electron spin; $x=Q^2/ 2M\nu$ is the Bjorken
variable; $g_1^A$ and $g_2^A$
are
the nuclear SSF; $F_2^A$ is the spin--independent
structure function of the target $A$; $R(x,Q^2)=\sigma_L (x,Q^2)/
\sigma_T (x,Q^2)$;
$\epsilon_1$ and $\epsilon_2$ are the energies of the incoming and
outgoing electrons.

In order
to extend to polarized DIS the usual convolution approach
adopted to treat the unpolarized DIS \cite{cps3},
let
us first consider the general case of inclusive scattering by spin
${1 \over 2}$ targets in impulse
approximation. For
the nuclear spin structure functions $g_1^A$ and $g_2^A$ one gets
\be
g_1^A(x,Q^2) & = & \sum_N \int
\, dE  \, d {\bf p} \, dz
\Bigg \{ { 1 \over z }g_1^N
\left( {x \over z},Q^2 \right)  \Bigg[ P_{||}^N({\bf p},E) \Bigg.
\Bigg. \nonumber \\
& & - {|{\bf p}| \over M} \left( {\nu \over |{\bf q}|}
 - {|{\bf p}| \cos \alpha \over E_p+M}
 \right){\cal P}^N ({\bf p},E)\Bigg]
-  {Q^2 \over |{\bf q}|^2}{1 \over M}{\cal L}^N \Bigg \} \nonumber \\
& & \delta \left( z - {p \cdot q \over M \nu} \right)
\label{qg1}
\ee
and
\be
g_2^A(x,Q^2) & = & \sum_N \int \,
dE  \, d {\bf p} \, dz
{1 \over M}
\Bigg \{ \Bigg [{ 1 \over z }g_1^N
\left( {x \over z},Q^2 \right) {\nu \over |{\bf q}|} |{\bf p}| {\cal P}^N
({\bf p},E)
 \Bigg. \Bigg. \nonumber \\
& &  + {1 \over z^2}  g_2^N \left( {x \over z},Q^2 \right)
\left(E_p  P_{||}^N({\bf p},E) - {|{\bf p}|^2 \cos \alpha \over M+E_P}
{\cal P}^N({\bf p},E) \right)
\Bigg ] \Bigg. \nonumber \\
& & - { \nu^2 \over |{\bf q}|^2} {\cal L}^N \Bigg \}
\delta \left( z - {p \cdot q \over M \nu} \right)
\label{qg2}
\ee
with
\be
{\cal L}^N= \left[ {1\over z} g_1^N \left( {x \over z},Q^2 \right)
{\cal H}_1^N + {|{\bf q}| \over \nu}{1\over z^2}  g_2^N
\left( {x \over z},Q^2 \right)
{\cal H}_2^N \right] \label{lst}
\ee
\be
{\cal H}_1^N = \left( {1 \over 2} \right){3 \cos^2 \alpha -1 \over \cos \alpha}
\left[ {|{\bf p}|^2 \over (E_p+M)}{\cal P}^N({\bf p},E)+
M { P_{\bot}^N({\bf p},E)\over \sin \alpha} \right]
\ee
\be
{\cal H}_2^N & = & |{\bf  p}| \left[ {\cal P}^N({\bf p},E)-
 { P_{\bot}^N({\bf p},E)\over \sin \alpha} \right] \nonumber \\
& & -{\nu \over 2 |{\bf q}|} {3 \cos^2 \alpha -1 \over \cos \alpha}
\left[ {|{\bf p}|^2 \over (E_p+M)}{\cal P}^N({\bf p},E)-E_p
 { P_{\bot}^N({\bf p},E)\over \sin \alpha} \right]~~.
\ee
In the above equations the sums extend over all nucleons $N$;
$p\equiv(p^0,{\bf p})$ is the four--momentum of the bound nucleon,
with $p^0=M_A-\left[(E-M+M_A)^2+|{\bf p}|^2 \right]^{1 \over 2}$;
$E$ is the nucleon removal energy;
$E_p=\left[
M^2 + |{\bf p}|^2 \right]^{1 \over 2}$;
$\cos \alpha = {\bf p} \cdot {\bf q} /|{\bf p}||{\bf q}|$. The quantities
$P_{||}^N({\bf p},E)$, $P_{\bot}^N({\bf p},E)$,
${\cal P}^N({\bf p},E)$ are defined as follows
\cite{cps1}:
\be
P_{||}^N ({\bf p},E)& = & P^N_{{1 \over 2}{1\over 2}M}({\bf p},E)-
P^N_{-{1 \over 2}-{1\over 2}M}({\bf p},E), \label{par} \\
P_{\bot}^N({\bf p},E) & = & 2 P^N_{{1 \over 2}-{1\over 2}M}({\bf p},E)
e^{i \phi}, \label{ort} \\
{\cal P}^N ({\bf p},E)& = & \sin \alpha P_{\bot}^N({\bf p},E)+
\cos \alpha P_{||}^N({\bf p},E) \label{Pi}, \label{cors}
\ee
where $\phi$ is the polar angle, and
\be
P_{\sigma \sigma'M}^N ({\bf p},E) & = & \sum_f[\langle
\psi^f_{A-1};N,{\bf p},\sigma'|\psi_{J,M} \rangle]^*  [\langle
\psi^f_{A-1};N,{\bf p},\sigma|\psi_{J,M} \rangle] \nonumber \\
& & \delta(E-E_{A-1}^f+E_A)~~. \label{spet}
\ee
is the spin dependent spectral function.
Of particular relevance are the ``up" and ``down" spectral functions
$P^N_{{1 \over 2}{1 \over 2}{1 \over 2}}$ and $P^N_{-{1 \over 2}-{1 \over 2}
{1 \over 2}}$, respectively, for they determine the nucleon polarizations, viz.
the quantities
\be
P_N^{(+)} & = & \int P^N_{{1 \over 2}{1 \over 2}{1 \over 2}} ({\bf p},E)
d {\bf p} dE
\label{up} \\
P_N^{(-)} & = & \int P^N_{-{1 \over 2}-{1 \over 2}{1 \over 2}} ({\bf p},E)
d {\bf p} dE~~. \label{down}
\ee
representing the probability
to have a proton (neutron) with spin parallel (+) or antiparallel (--)
to $^3$He spin.
Using in Eqs.\ (\ref{qg1}) e (\ref{qg2}) the proper nucleon SSF $g_{1(2)}^N$,
the nuclear SSF
$g_{1(2)}^A$ can be evaluated in the quasielastic, inelastic and DIS regions.

In the Bjorken
limit  $(\nu /
| {\bf q} | \rightarrow 1$, $Q^2/| {\bf q} |^2 \rightarrow 0)$,
the nuclear asymmetry reduces to the following expression
\be
A_{||}=2x{g_1^A(x) \over F_2^A(x)} \label{asv}
\ee
where
\be
g_1^A(x)  =  \sum_N \int _x ^A dz
{1 \over z} g_1^N \left( {x \over z} \right)
G^N_1(z)~~, \label{fin}
\ee
with the spin dependent light cone momentum distribution $G^N_1(z)$ given by
\be
G_1^N(z)  =  \int dE\, \int d {\bf p}
\bigg \{  P_{||}^N( {\bf p},E )- \left[ 1 -
{p_{||} \over E_p + M} \right] {|{\bf p}| \over M}
{\cal P}^N({\bf p},E) \bigg\}
\delta \left(z - {p^+ \over M} \right)~~, \label{lux}
\ee
$p^+=p^0-p_{||}$ being the nucleon light cone momentum.
It should be pointed out that if
the term proportional to ${\cal P}^N$
in Eq.\ (\ref{lux}) is disregarded, $G_1^N(z)$ reduces to the difference
between the ``up", $f_N^{(+)}$, and ``down", $f_N^{(-)}$, light--cone momentum
distributions, viz. $G_1^N(z)=f_N^{(+)}(z)-f_N^{(-)}(z)$, where
\be
f_N^{(+)} (z) & = & \int dE \int
 d {\bf p} \, P^N_{{1 \over 2}{1 \over 2}{1 \over 2}} ({\bf p},E)
\, \delta \left(z - {p^+ \over M} \right)
\label{zup} \\
f_N^{(-)} (z) & = & \int dE \int
 d {\bf p} \,  P^N_{-{1 \over 2}-{1 \over 2}{1 \over 2}} ({\bf p},E) \,
\delta \left(z - {p^+ \over M} \right)~~. \label{zdown}
\ee
Three models for $^3$He asymmetry, in order of increasing
complexity, have been considered, viz.:
\\
1) {\it No nuclear effects}. One assumes that
\be
g_1^3(x,Q^2) & = & g_1^n(x,Q^2) \label{gmod1} \\
A_{\vec {^3{\rm He}}} & = & f_n A_{\vec n} \label{mod1}
\ee
where $A_{\vec n}(x,Q^2)$
is the neutron asymmetry and
$f_n=F_2^n(x,Q^2)/[2F_2^p(x,Q^2) + F_2^n(x,Q^2)]$ the neutron
dilution factor. Such a picture
is equivalent to considering polarized electron scattering off
$^3{\vec{\rm He}}$ described as a
pure symmetric $S$ wave
disregarding, moreover, Fermi motion and binding effects.
\\
2) {\it Proton contribution within realistic wave function of $^3$He.}
Besides the $S$ wave, the three body wave function contains a
percentage of $S'$ and $D$ waves,
$P_{S'}$ and $P_D$,
which are responsible for a proton
contribution to the polarization of $\vec{^3{\rm He}}$. The amount of such a
contribution can be determined by
calculating the nucleon polarizations $P_N^{(\pm)}$
using the ground state wave function.
In a pure $S$ wave state $P_n^{(+)}=1$, $P_n^{(-)}=0$ and
$P_p^{(+)}=P_p^{(-)}={1\over2}$, whereas for a three--body wave function
containing $S$, $S'$ and $D$ waves, one has
\cite{fri,kap}
\be
P_n^{(\pm)}={1\over 2} \pm {1 \over 2} \mp \Delta, \label{pn}
\ee
\be
P_p^{(\pm)}={1 \over 2}
\mp \Delta'~~, \label{pp}
\ee
where $\Delta={1 \over 3} [P_{S'}+2P_D]$ and
$\Delta'={1 \over 6}[P_D-P_{S'}]$.
{}From
world calculations on the three body system one obtains,
in correspondence of the experimental value of the
binding energy of $^3$He,
$\Delta=0.07 \pm 0.01$ and $\Delta'=
0.014\pm 0.002$ \cite{fri}.
It is also useful to express $P_{N}^{(\pm)}$ in terms of the spin dependent
light--cone momentum distributions [Eqs. (\ref{zup})
and (\ref{zdown})], namely
\be
P_N^{(\pm)}=\int f_N^{(\pm)}(z)dz~~. \label{pup}
\ee
Thus if the $S'$ and $D$ waves are
considered and
Fermi motion and binding effects
are disregarded, one can write
\be
g_1^3(x,Q^2) & = & 2p_p g_1^p(x,Q^2) + p_n g_1^n(x,Q^2)\label{gmod2} \\
A_{\vec {^3{\rm He}}}&  = & 2 f_p p_p A_{\vec p} + f_n p_n A_{\vec n}
\label{mod2}
\ee
where $f_{p(n)}(x,Q^2)=
F_2^{p(n)}(x,Q^2)/[2F_2^p(x,Q^2)+F_2^n(x,Q^2)]$
is the proton (neutron) dilution factor,
$A_{{\vec p}({\vec n})}(x,Q^2)=2x g_1^{p(n)}(x,Q^2)/F_2^{p(n)}(x,Q^2)$
is the proton (neutron) asymmetry
and $p_{p(n)}$ are the effective nucleon polarizations
\be
p_p & = & P_p^{(+)}-P_p^{(-)}=-0.028{\pm} 0.004
\label{polp} \\
p_n & = & P_n^{(+)}-P_n^{(-)}=0.86 {\pm} 0.02 \label{pol}
\ee
The above values correspond to Eqs. (\ref{pn}) and (\ref{pp}), while
Eq. (\ref{pup}) with our spin dependent spectral function yields
$p_p=-0.030$ and $p_n=0.88$.
\\
3) {\it Proton contribution within the convolution approach.}
The asymmetry is given by Eqs. (\ref{asv}), (\ref{fin})
 and (\ref{lux}) in the Bjorken limit
and by Eqs. (\ref{as})--(6) at finite values of $Q^2$.

Calculations have been performed using for the
nucleon SSF $g_1^N$
the one proposed in Ref.
\cite{scha} and for the
effective nucleon polarization $p_{p(n)}$,
the values given by Eqs.\ (\ref{polp})
and (\ref{pol}).

In Fig. 1 the neutron and proton
spin--dependent light cone momentum distributions in $^3$He,
[Eq. (\ref{lux})] are presented.
Because of the small contribution from ${\cal P}^N$ they are almost enterily
given by the difference between the ``up" and ``down"
light cone momentum distributions $f^{(\pm)}_N$
[Eqs. (\ref{zup}) and (\ref{zdown})], which are shown
in Fig. 2.
The Bjorken limit of the ${^3\vec{\rm He}}$ asymmetry [Eq.\ (\ref{asv})]
is presented in Fig. 3(a).
It can be seen that the proton contribution is very important
at $x \geq 0.3$, and therefore it hinders in principle
the extraction of the neutron structure
function.
In order to understand how much of the proton contribution
 to the asymmetry is given by
$S'$ and $D$ waves, by Fermi motion and
by binding effects,
the
asymmetry
predicted by the convolution approach is compared in Fig. 3(b)
with the predictions of the
models 1 and 2 [Eqs. (\ref{gmod1})--(\ref{mod2})].
It can be
seen that
the asymmetry which totally lacks of nuclear effects
(binding and Fermi motion as well as $S'$ and $D$ waves),
strongly differs from the ones which
include these
effects; however it can also be seen that
at least for $x \leq 0.9$
nuclear effects can reliably be taken care of
by Eq.\ (\ref{mod2}), i.e. by considering
only the effective nucleon
polarization induced by $S'$ and $D$ waves.
If so, the neutron asymmetry $\tilde{A_{\vec n}}$
can be obtained from the $^3$He
asymmetry using Eq. (\ref{mod2}), viz.
\be
\tilde{A_{\vec n}}(x)= {1 \over f_n p_n}
\left[ A_{\vec {^3He}}(x)-2 p_p f_p A_{\vec p}(x) \right]~~. \label{at}
\ee
In Fig. 4(a), $\tilde{A_{\vec n}}$ calculated from
Eq. (\ref{at}) using the convolution formula
for $A_{\vec {^3He}}(x)$ is compared with the free neutron asymmetry;
it can be seen
that the two quantities are very close to each other, differing,
because of
binding and Fermi motion effects, by at most $4\%$ for $x \leq 0.9$.
The same comparison for the SSF is presented in Fig. 4(b), where
the free neutron SSF
is compared with the quantity:
\be
\tilde g_1^n(x)= {F_2^n(x){\tilde{A_{\vec n}}}(x) \over 2x}~~. \label{g1art}
\ee
It has been shown in Ref. \cite{pg} that
the small effects from binding and Fermi motion
can be understood
by expanding ${1 \over z} g_1^N \left( {x \over z} \right)$
in Eq.\ (\ref{fin}) around $z=1$ and
by disregarding the term proportional to ${\cal P}^N$
in Eq.\ (\ref{lux}); by this way
the correctness of Eq.\ (\ref{gmod2})
can be easily justified.

It appears therefore that the only relevant nuclear effects
in inclusive DIS of polarized
electrons off polarized $^3$He in the Bjorken limit are those related
to the proton and neutron effective polarizations arising
from $S'$ and $D$ waves; such a result, moreover, does not seem to
crucially depend
upon the form of $g_1^N$.

 In order to perform a significant comparison between
theoretical predictions and
experimental data,
one has to investigate the $Q^2$ dependence of $g_1^3(x,Q^2)$
by evaluating Eq. (\ref{qg1}). Taking \cite{jaf}
\be
g_2^N(x,Q^2)=-g_1^N(x,Q^2)+\int_x^1 dy {1\over y} g_1^N(y,Q^2) \label{g2}
\ee
and
assuming the kinematics of Ref. \cite{slac2},
we found that $g_1^3(x,Q^2)$ and $g_1^3(x)$ differ by at most
$15\%$. These are preliminary results and it is not the aim of this talk
to discuss their possible implications on the Bjorken sum rule.

Future experiments are planning the direct measurement of $g_2^n$ and $g_2^3$;
the first quantity \cite{jaf} is shown in Fig. 5(a) whereas
$g_2^3$ calculated taking
into account the proton and neutron effective polarizations,
i.e. using the expression resulting from the series expansion of
${1 \over z^2} g_2^N \left( {x \over z} \right)$ around $z=1$
in Eq. (\ref{qg2}), is presented in Fig. 5(b).

A detailed analysis of the transverse asymmetry, as well as of the
$Q^2$ dependent
effects, is in progress and will be reported elsewhere.
\vskip 15mm

\eject
\centerline{\bf \Large Captions}
\begin{quotation}
FIG. 1. (a) The nucleon spin--dependent light--cone momentum
distributions $G^N_1(z)$
in $^3$He [Eq.\ (\protect \ref{lux})] for the neutron (a)
and the proton (b), respectively.
\end{quotation}

\begin{quotation}
FIG. 2. (a) The ``up" and ``down"
light--cone momentum distributions $f_N^{(+)}(z)$ [Eq.\ (\protect \ref{zup})]
(full) and $f_N^{(-)}(z)$ [Eq.\ (\protect \ref{zdown})] (dashed) for the
neutron
(a) and the proton (b) in $^3$He.
\end {quotation}

\begin{quotation}
FIG. 3. (a) The $^3$He asymmetry [Eq.\ (\protect \ref{as})]
calculated within the convolution
approach [Eq.\ (\protect \ref{fin})](full). Also shown are the
neutron (short--dashed) and proton (long--dashed) contributions.
(b) The $^3$He asymmetry
calculated within different nuclear
models. Dotted line: no nuclear effects [Eq.\ (\protect \ref{mod1})];
short--dashed line: $S'$ and $D$ waves
of $^3$He taken into account [Eq.\ (\protect \ref{mod2})];
long--dashed line: $S'$ and $D$ waves
of $^3$He taken into account plus Fermi motion effects;
full line: $S'$ and $D$ waves
of $^3$He taken into account plus
Fermi motion and
binding effects [after \protect \cite{pg}].
\end {quotation}

\begin{quotation}
FIG. 4. (a) The free neutron asymmetry (dots) compared
with the neutron asymmetry given by Eq. (\protect \ref{at})(full).
(b) The same as in (a) but for the SSF $g_1^n$.
The dotted line represents the free neutron structure function $g_1^n$,
whereas the
full line represents the neutron structure function given by
Eq.\ (\protect \ref{g1art}). The difference between
the two curves
is due to Fermi motion and binding effects [after \protect \cite{pg}].
\end{quotation}

\begin{quotation}
FIG. 5. (a) The free neutron structure function $g_2^n$
\protect \cite{jaf};(b) the $^3$He structure function $g_2^3$
calculated taking into account the effective proton and neutron
polarizations (full);
the neutron and proton contribution are represented by the short--dashed
and long--dashed curves, respectively.
\end{quotation}

\eject

\end{document}